\documentclass[%
 reprint,
superscriptaddress,
 amsmath,amssymb,
 aps,
]{revtex4-1}
\usepackage{graphicx}
\usepackage{bm}
\usepackage{epsf}
\usepackage{rotating}
\usepackage{epsfig,graphics,rotate,color}
\usepackage{wrapfig}
\usepackage{amssymb}
\usepackage{amsmath}
\usepackage{amsfonts}
\usepackage{array,hhline,dcolumn}%
\providecommand{\U}[1]{\protect\rule{.1in}{.1in}}
\begin{document}

\title{Direct Measurement of Backgrounds using Reactor-Off
  Data in Double Chooz}

\begin{abstract}

Double Chooz is unique among modern reactor-based neutrino experiments studying  $\bar \nu_e$ disappearance in that data can be collected with all reactors off.    
In this paper, we present data from 7.53 days of reactor-off running.   
Applying the same selection criteria as used in the Double Chooz reactor-on oscillation analysis, a measured background rate of 1.0$\pm$0.4 events/day is obtained. 
The background model for accidentals, cosmogenic $\beta$-$n$-emitting isotopes, fast neutrons from cosmic muons, and stopped-$\mu$ decays  used in the oscillation analysis is 
demonstrated to be correct within the uncertainties.
Kinematic distributions of the events, which are dominantly cosmic-ray-produced correlated-background events, are provided.   
The background rates are scaled to the shielding depths of two other reactor-based oscillation experiments, Daya Bay and RENO.

\end{abstract}

\newcommand{\Aachen}{III. Physikalisches Institut, RWTH Aachen 
University, 52056 Aachen, Germany}
\newcommand{\Alabama}{Department of Physics and Astronomy, University of 
Alabama, Tuscaloosa, Alabama 35487, USA}
\newcommand{\Argonne}{Argonne National Laboratory, Argonne, Illinois 
60439, USA}
\newcommand{\APC}{APC, AstroParticule et Cosmologie, Universit\'{e} Paris 
Diderot, CNRS/IN2P3, CEA/IRFU, Observatoire de Paris, Sorbonne Paris 
Cit\'{e}, 75205 Paris Cedex 13, France}
\newcommand{\CBPF}{Centro Brasileiro de Pesquisas F\'{i}sicas, Rio de 
Janeiro, RJ, cep 22290-180, Brazil}
\newcommand{\Chicago}{The Enrico Fermi Institute, The University of 
Chicago, Chicago, IL 60637, USA}
\newcommand{\CIEMAT}{Centro de Investigaciones Energ\'{e}ticas, 
Medioambientales y Tecnol\'{o}gicas, CIEMAT, E-28040, Madrid, Spain}
\newcommand{\Columbia}{Columbia University; New York, NY 10027, USA}
\newcommand{\Davis}{University of California, Davis, CA-95616-8677, USA}
\newcommand{\Drexel}{Physics Department, Drexel University, Philadelphia, 
Pennsylvania 19104, USA}
\newcommand{\Hamburg}{Institut f\"{u}r Experimentalphysik, 
Universit\"{a}t Hamburg, 22761 Hamburg, Germany}
\newcommand{\Hiroshima}{Hiroshima Institute of Technology, Hiroshima, 
731-5193, Japan}
\newcommand{\IIT}{Department of Physics, Illinois Institute of 
Technology, Chicago, Illinois 60616, USA}
\newcommand{\INR}{Institute of Nuclear Research of the Russian Aacademy 
of Science, Russia}
\newcommand{\CEA}{Commissariat \`{a} l'Energie Atomique et aux Energies 
Alternatives, Centre de Saclay, IRFU, 91191 Gif-sur-Yvette, France}
\newcommand{\Livermore}{Lawrence Livermore National Laboratory, 
Livermore, CA 94550, USA}
\newcommand{\Kansas}{Department of Physics, Kansas State University, 
Manhattan, Kansas 66506, USA}
\newcommand{\Kobe}{Department of Physics, Kobe University, Kobe, 
657-8501, Japan}
\newcommand{\Kurchatov}{NRC Kurchatov Institute, 123182 Moscow, Russia}
\newcommand{\MIT}{Massachusetts Institute of Technology; Cambridge, MA 
02139, USA}
\newcommand{\MaxPlanck}{Max-Planck-Institut f\"{u}r Kernphysik, 69117 
Heidelberg, Germany}
\newcommand{\Niigata}{Department of Physics, Niigata University, Niigata, 
950-2181, Japan}
\newcommand{\NotreDame}{University of Notre Dame, Notre Dame, IN 46556-
5670, USA}
\newcommand{\IPHC}{IPHC, Universit\'{e} de Strasbourg, CNRS/IN2P3, F-
67037 Strasbourg, France}
\newcommand{\SUBATECH}{SUBATECH, CNRS/IN2P3, Universit\'{e} de Nantes, 
Ecole des Mines de Nantes, F-44307 Nantes, France}
\newcommand{\Sussex}{Department of Physics and Astronomy, University of 
Sussex, Falmer, Brighton BN1 9QH, United Kingdom}
\newcommand{\Tennessee}{Department of Physics and Astronomy, University 
of Tennessee, Knoxville, Tennessee 37996, USA}
\newcommand{\TohokuUni}{Research Center for Neutrino Science, Tohoku 
University, Sendai 980-8578, Japan}
\newcommand{\TohokuGakuin}{Tohoku Gakuin University, Sendai, 981-3193, 
Japan}
\newcommand{\TokyoInst}{Department of Physics, Tokyo Institute of 
Technology, Tokyo, 152-8551, Japan  }
\newcommand{\TokyoMet}{Department of Physics, Tokyo Metropolitan 
University, Tokyo, 192-0397, Japan}
\newcommand{\Muenchen}{Physik Department, Technische Universit\"{a}t 
M\"{u}nchen, 85747 Garching, Germany}
\newcommand{\Tubingen}{Kepler Center for Astro and Particle Physics, 
Universit\"{a}t T\"{u}bingen, 72076, T\"{u}bingen, Germany}
\newcommand{\UFABC}{Universidade Federal do ABC, UFABC, Sao Paulo, Santo 
Andr\'{e}, SP, Brazil}
\newcommand{\UNICAMP}{Universidade Estadual de Campinas-UNICAMP, 
Campinas, SP, Brazil}
\newcommand{\Aviette}{Laboratoire Neutrino de Champagne Ardenne, domaine 
d'Aviette, 08600 Rancennes, France}
\newcommand{\vtech}{Center for Neutrino Physics, 
Virginia Tech, Blacksburg, VA}
\newcommand{\deceased}{Deceased.}

\affiliation{\Aachen}
\affiliation{\Alabama}
\affiliation{\Argonne}
\affiliation{\APC}
\affiliation{\CBPF}
\affiliation{\Chicago}
\affiliation{\CIEMAT}
\affiliation{\Columbia}
\affiliation{\Davis}
\affiliation{\Drexel}
\affiliation{\Hamburg}
\affiliation{\Hiroshima}
\affiliation{\IIT}
\affiliation{\INR}
\affiliation{\CEA}
\affiliation{\Livermore}
\affiliation{\Kansas}
\affiliation{\Kobe}
\affiliation{\Kurchatov}
\affiliation{\MIT}
\affiliation{\MaxPlanck}
\affiliation{\Niigata}
\affiliation{\NotreDame}
\affiliation{\IPHC}
\affiliation{\SUBATECH}
\affiliation{\Muenchen}
\affiliation{\Tennessee}
\affiliation{\TohokuUni}
\affiliation{\TohokuGakuin}
\affiliation{\TokyoInst}
\affiliation{\TokyoMet}
\affiliation{\Tubingen}
\affiliation{\UFABC}
\affiliation{\UNICAMP}
\affiliation{\vtech}

\author{Y.~Abe}
\affiliation{\TokyoInst}

\author{C.~Aberle}
\affiliation{\MaxPlanck}

\author{J.C.~dos Anjos}
\affiliation{\CBPF}

\author{J.C.~Barriere}
\affiliation{\CEA}

\author{M.~Bergevin}
\affiliation{\Davis}

\author{A.~Bernstein}
\affiliation{\Livermore}

\author{T.J.C.~Bezerra}
\affiliation{\TohokuUni}

\author{L.~Bezrukhov}
\affiliation{\INR}

\author{E.~Blucher}
\affiliation{\Chicago}

\author{N.S.~Bowden}
\affiliation{\Livermore}

% ...

\collaboration{Double Chooz Collaboration}


\maketitle

With the discovery of the last mixing angle of the three-neutrino mixing matrix
\cite{dc2011,dc2012,DB,RENO,T2K,MINOS}, neutrino oscillation experiments
entered the precision era.  The next goal is precision
studies of the three-active-neutrino model,  
including searches for $CP$-violation or non-unitarity \cite{model}.   The transition from searches to 
precision measurements necessitates a 
higher standard for understanding 
backgrounds to oscillation analyses.  

Among ongoing reactor-based oscillation experiments,
Double Chooz (DC) is unique in obtaining a ``reactor-off''
data set when the  two cores of the Chooz site are both brought 
down for maintenance.  
The Daya Bay \cite{DB} and 
RENO \cite{RENO} experiments are each located at complexes with six cores.
Consequently, they are unlikely to obtain data with all cores off.
The CHOOZ experiment reported reactor-off running  \cite{Chooz},  but 
with varying scintillator stability and higher accidentals rate and threshold
than in DC.
We present here the results of 7.53 days of DC reactor-off
running, collected in 2011 and 2012.
This data set demonstrates the validity 
of the background predictions for present and
future $\theta_{13}$ experiments.  


The primary goal of DC is measurement of the neutrino oscillation parameter $\theta_{13}$ through $\bar \nu_e$ disappearance.   
The design of the Daya Bay and RENO detectors is similar to that of  DC \cite{dc2012}.
All three experiments use the inverse beta decay (IBD) interaction ($\bar \nu_e + p \rightarrow e^+ + n$) in liquid scintillator.    
This interaction is identified by a correlated pair of signals, the first consistent with a positron and the second consistent with a $n$-capture.

The DC far detector is positioned 1050 m from the two 4.25 GW$_{th}$ (thermal power)
cores of the Chooz Nuclear Power Plant.   It 
consists of four concentric
cylindrical regions, with centered
chimneys for filling and
insertion of calibration sources.
The innermost cylinder is the ``Neutrino Target'' (NT), a 10 m$^3$
volume of gadolinium-doped
liquid scintillator.   The acrylic NT cylinder is surrounded by a 55 cm thick ``$\gamma$
Catcher'' (GC) consisting of Gd-free scintillator.   The acrylic cylinder
of the GC is immersed in a 105 cm thick nonscintillating oil
``buffer region''  containing 390 10-inch photomultiplier tubes (PMT).   These three
cylinders, collectively called the ``inner detector'' (ID),  
are contained in a stainless steel vessel which is encompassed
by a 50 cm thick liquid scintillator region forming the ``Inner
Veto'' (IV).   The IV is
surrounded by 15 cm of demagnetized steel, followed by
rock.  Above this system 
is the ``Outer Veto'' (OV),   consisting of segmented scintillator modules
for muon tracking.

The detector is shielded
from cosmic rays by a 300 meters water equivalent (m.w.e.)\ rock overburden, in a hill topology.  The 
dominant backgrounds  in the reactor neutrino experiments are:
spallation products, particularly $^9$Li and $^8$He, produced by cosmic muons interacting in oil, which emit a $n$ immediately following the $\beta$-decay process; stopping muons; and fast neutrons produced by muons in the surrounding rock.
In this Letter, we refer to the first as ``$\beta$-$n$ backgrounds,'' while the latter two are collectively called ``$\mu$/fast-$n$'' backgrounds.  
These are directly measured by reactor-off running.     
The DC overburden being similar to those of Daya Bay and RENO, these results can be applied to those experiments with modest scaling for depth variations.


A direct measurement of the backgrounds in the DC oscillation analyses is performed by applying the same $\bar{\nu}_e$ selection criteria as in Refs.\ \cite{dc2011} and \cite{dc2012} to the reactor-off data sample.  
A minimal set of selection cuts was applied in 
\cite{dc2011} (``DCI selection''). Two extra cuts were added in 
\cite{dc2012}  (``DCII selection'') to reduce background contamination in the $\bar{\nu}_e$ candidate sample. 
The results presented here apply to both the DCI and DCII selections, comparing the reactor-off data with expectations from the published reactor-on oscillation analyses~\cite{dc2012}.

Candidates are extracted from a sample of triggers (``singles'') above 0.5 MeV that are neither tagged as a background known as ``light noise,'' nor vetoed by the 1 ms muon veto ($\mu$ veto) \cite{dc2012}. 
The DCI selection then applies four cuts to the prompt ($e^+$) and delayed ($n$) IBD signals: 
1) time difference: $2~\mu$s $<$ $\Delta t_{{\rm prompt}/n}$ $<$ $100~\mu$s; 
2) prompt trigger: $0.7~$
$ < E_{\rm prompt} < 12.2~$MeV; 
3) delayed trigger: $6.0~$
$ < E_{n} < 12.0~$MeV;
4)  multiplicity: no additional valid triggers from $100~\mu$s preceding the prompt signal to $400~\mu$s after it. 
The DCII selection further rejects candidates according to two more conditions: 
5) cosmogenic $\beta$-$n$ background reduction: candidates within a 0.5 s window after a muon depositing high energy ($>$600 MeV) crosses the ID (``showering-$\mu$ veto''); 
6) $\mu$/fast-$n$ background reduction: candidates whose prompt signal is coincident with an OV signal (OV veto).

During the reactor-off period, the total and showering muon rates (ID
only) were 46 and 0.10 s$^{-1}$, respectively, 
consistent with those during the reactor-on period to 4\% \cite{dc2012}. 
By applying the $\mu$ veto without and with the additional DCII showering-$\mu$ veto, 7.19 and 6.84 live days, respectively, are obtained. 
Within these times, a singles rate of 11.01 s$^{-1}$ is measured,
again consistent, within 4\%, with that during the
reactor-on period. Hence, the same accidental background level is
expected for DCI and DCII.

Table~\ref{tab:expBkg} shows the estimated background and observed reactor-off event rates for both the DCI and DCII selections.
In all cases, the background rate estimation relies on data published in~\cite{dc2012}.
The accidental rate uncertainties quoted include an additional effect of day-to-day variations, negligible in~\cite{dc2012}.
For the DCII selection, the $^9$Li rate corresponds to the value used as an input for the oscillation fit, which is consistent with the fit output,
and the $\mu$/fast-$n$ rate is smaller than that reported in~\cite{dc2012} since OV duty-cycle was 100\% during the reactor-off period.

\begin{table}[tb]
\caption{
Background rate estimates~\cite{dc2012}, in events/day, for the reactor-off data sample, compared to observation, for the two selections described in the text.
\label{tab:expBkg}}
\begin{tabular}{cccc|c|c}
\hline \hline 
Rate         &  $\beta$-$n$  & Accidental  & $\mu$/fast $n$ & Total & Total \\
(day$^{-1}$) &               &             &                &  Est. & Obs. \\ \hline\hline
DCI    &  2.10$\pm$0.57  &  0.35$\pm$0.02  &  0.93$\pm$0.26 &
3.4$\pm$0.6 & 2.7$\pm$0.6\\
DCII   &  1.25$\pm$0.54  &  0.26$\pm$0.02  &  0.44$\pm$0.20 &
2.0$\pm$0.6& 1.0$\pm$0.4 
\\
\hline \hline 
\end{tabular} 
\end{table}

In order to evaluate the residual neutrino spectrum in the reactor-off
period, a dedicated simulation has been performed with FISPACT
\cite{FISPACT}, an evolution code predicting the isotope inventory in
the reactor cores. The neutrino spectrum is then computed using the
BESTIOLE \cite{BESTIOLE} database. 
The resulting total number of expected neutrino interactions 
during the reactor-off period is 
2.01$\pm$0.80, 
which, when corrected for 
the live time ($\mu$ vetoes) and the detection efficiency computed
in \cite{dc2012}, yields an expected number of detected neutrino events of
1.49$\pm$0.60
(1.42$\pm$0.57) in the DCI (DCII) analysis.
The dominant contribution comes from long-half-life isotopes, 
so the time distribution of these events is expected to be essentially 
flat over the several-day reactor-off period.

\begin{figure}[tb]
\includegraphics[width=0.9\linewidth]{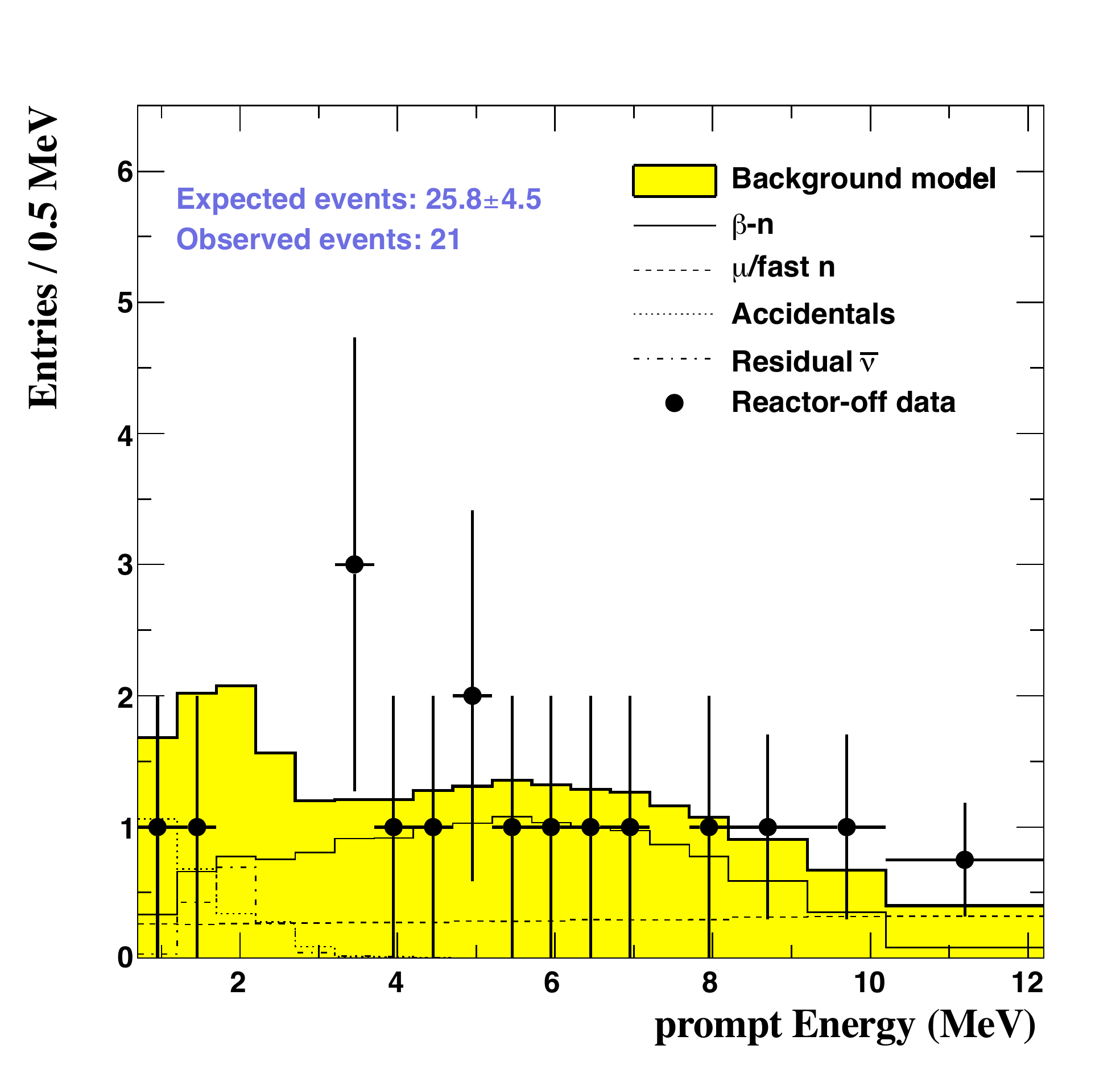} 
\includegraphics[width=0.9\linewidth]{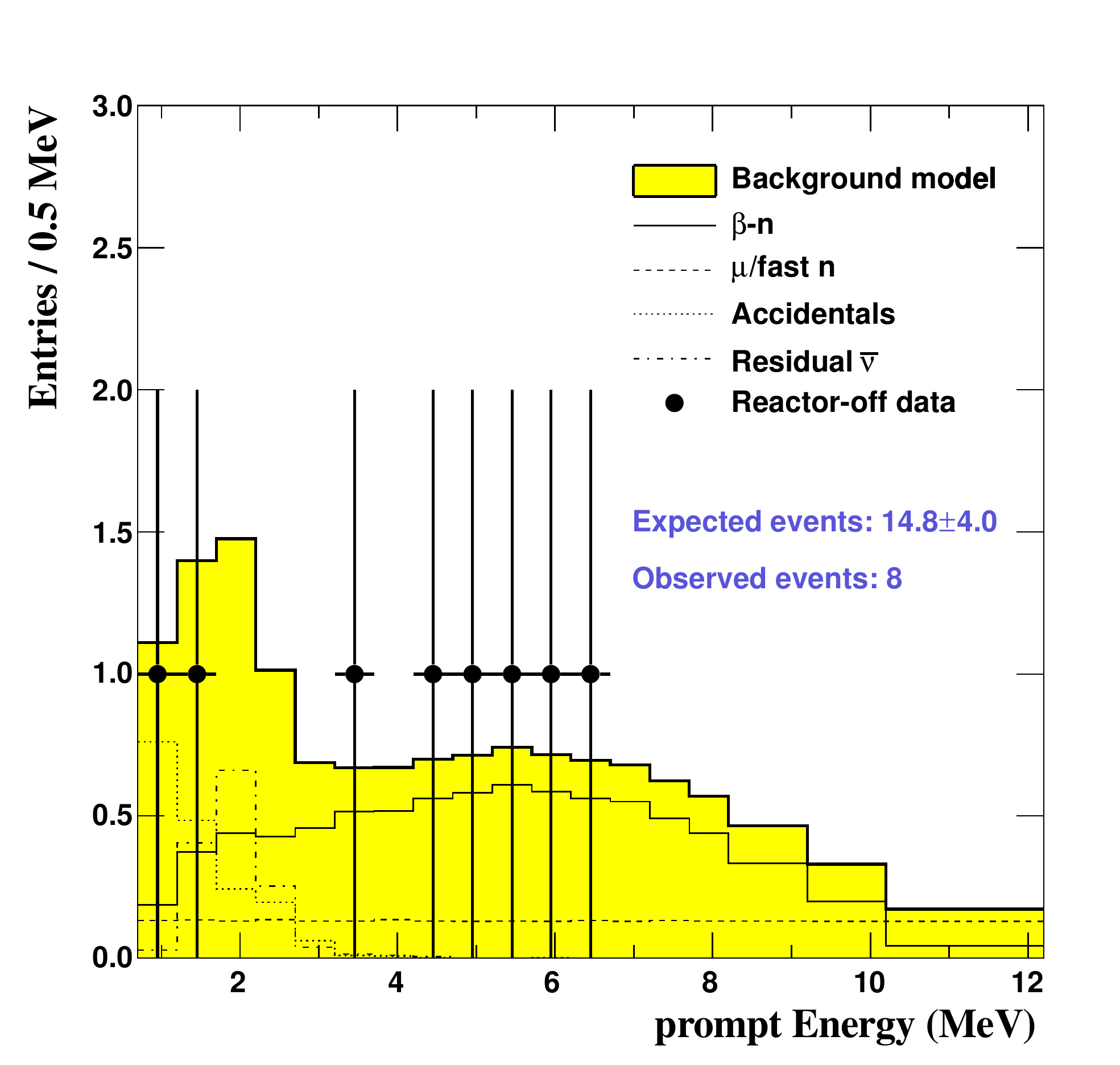} 
\caption{$\bar{\nu}_e$ candidates in the reactor-off data
  sample, with breakdown by components. Top and bottom figures show DCI and DCII selection results, respectively. Black points: data; histogram:
  background+$\bar{\nu}_e$ expectation. 
\label{fig:DCIBKG}}
\end{figure}

The application of the $\bar{\nu}_e$ selection cuts to the reactor-off
data sample yields 21 (8) $\bar{\nu}_e$ candidates in the
DCI (DCII) analysis. The DCII analysis vetoes five
events using the showering-$\mu$ veto ($\beta$-$n$-like events),
and another eight using the OV veto ($\mu$/fast-$n$-like events).
Figure \ref{fig:DCIBKG} shows the prompt energy
distribution of the candidates, superimposed on the expected spectra
of background events and residual neutrinos. 
Once the expected number of detected neutrinos is subtracted,
these numbers yield a measured total background of 2.7$\pm$0.6
events/day  (1.0$\pm$0.4 events/day) using DCI (DCII).
This result is consistent with the
background estimates, as shown in Table \ref{tab:expBkg},
confirming the reliability of 
the background model for the oscillation analysis.

The accidental background rate obtained in the reactor-off data
sample is 0.26$\pm$0.02 events/day, in perfect agreement with the prediction in Table~\ref{tab:expBkg}. Unlike other backgrounds, accidentals have no spatial correlation between the prompt and delayed signals. One event in the reactor-off sample with distance between the vertices $\Delta r \approx$3.5~m is clearly accidental-like.

Following the analysis presented in \cite{dc2012}, the
cosmogenic $\beta$-$n$  background rate can be determined from the time correlation
to the parent muon.  An exponential decay plus 
a constant background is fit to the time difference ($\Delta
t_{\mu\nu}$) distribution between muons and IBD
candidates. DCI selection plus the OV veto (to reduce 
$\mu$/fast-$n$ contamination) yields $1.7\pm0.9$
$\beta$-$n$-events/day.  The number remaining after  DCII
selection is
$1.1\pm0.8$
events/day.
The results are in good agreement with the $\Delta t_{\mu\nu}$ fit of
the reactor-on data, 
which indicated  $2.1\pm0.6$ ($1.3\pm 0.5$) events/day for DCI+OV
(DCII) selection~\cite{dc2012}. 
The five events tagged by the showering-$\mu$ veto correspond to a $\beta$-$n$
rate of 0.70$\pm$0.31 events/day, consistent with the value in~\cite{dc2012}: 0.89$\pm$0.10 events/day. 
 
A sample of stopping muons and fast neutrons is obtained by applying the OV veto (cut 6) to the candidates passing the DCI selection. 
Eight events are tagged by the OV in the range $E_{\rm prompt}$~=~0.7 to 12.2 MeV, while four 
are found between 12.2 and 30 MeV, where only $\mu$/fast-$n$ background is expected. 
Of these, ten events have $\Delta t<3~\mu$s, and their reconstructed vertices populate the region below the detector chimney. 
These are classified as stopping muons that decay. 
The remaining two candidates are farther from the chimney and have large $\Delta t$, as expected for fast-neutron events. 
The overall OV tagging rate for $E_{\rm prompt}<$~30~MeV in the reactor-off period is 1.67$\pm$0.48 events/day, in good agreement with that observed in the reactor-on data: 1.70$\pm$0.10 events/day. 
Both IV and OV tagging techniques \cite{dc2012} were applied to the reactor-off data, yielding results  consistent with those of the reactor-on analysis.

The rates of the IBD candidates originating from fast-$n$ 
(excluding stopped-$\mu$'s) and $\beta$-$n$ backgrounds 
can be scaled to other experimental sites, 
such as those of the Daya Bay and RENO detectors and the future DC near detector. 
As these  backgrounds are produced by muons, the first step is scaling the muon flux ($\Phi_\mu$) and mean energy ($\langle E_\mu\rangle$).
IBD rates from fast-$n$ and $\beta$-$n$ isotope production can then be computed.

The muon flux (in $\mu$/cm$^2$/s) at the DC far site is estimated using two independent  methods: 
the total measured muon rate ($\mu$/s) divided by either 1) the effective detector area, or
2)  the detector volume,  then  multiplied by average path length within the volume.
The two methods yield consistent results and are in agreement with a simulation using the MUSIC/MUSUN  code~\cite{Music}, which includes a detailed description of the  overburden topology. 
The results also agree with measurements by the CHOOZ experiment~\cite{Chooz}, once the definition of the effective area is correctly taken into account.
An average of 
estimates 1) and 2) is taken as the DC far flux,  with an error estimated from the difference between measurement and simulation.
A MUSIC/MUSUN simulation also yields the average  muon energy at the DC far site.
The values are summarized in Table~\ref{tab:scaleinputs}, including
measured rates of fast-$n$ and $\beta$-$n$ backgrounds.
The fast-$n$ rate was computed as in~\cite{dc2012} for the reactor-on
data sample, both using the OV veto (DCII) on the subsample where the
OV was fully operational, and on the whole sample excluding this cut (DCI).

\begin{table}[tb]
\caption{Values for the relevant quantities at the
DC far site, used as input for
scaling backgrounds with depth.
\label{tab:scaleinputs}}
\begin{tabular}{ll}
\hline\hline
Muon flux $\Phi_\mu^{DC}$ & 
0.72 $\pm$ 0.04 m$^{-2}$s$^{-1}$ \\
Mean muon energy $\langle E_\mu^{DC} \rangle$ &
63.7 $\pm$ 0.8 GeV \\
Fast-$n$ background rate  & 
   0.33 $\pm$ 0.16 d$^{-1}$ DCI \\
&  0.23 $\pm$ 0.18 d$^{-1}$ DCII \\
{$\beta$-$n$} background rate  & 
     1.7 $\pm$ 0.9 d$^{-1}$ DCI + OV \\
&    1.1 $\pm$ 0.8 d$^{-1}$ DCII  \\
\hline\hline
\end{tabular}
\end{table}

The measured muon flux was scaled following two different empirical methods~\cite{Reichenbacher,Bugaev}.
Both are applicable for shallow depths and provide consistent results.
Such methods assume a flat overburden. The shape of the overburden
affects the overall rate, but has only a minor impact on the evolution
of the rate with 
depth.
As a realistic evaluation of the effect, we find the difference between the rates for a flat overburden and the hill profile at the DC far site to be 11\%.

The mean muon energy 
was calculated at various depths using the MUSIC/MUSUN simulation code.
We take the uncertainty on these values due to  
overburden shape  to be 3.6\%: 
this comes from our calculations of the mean muon energies at a depth of 300 m.w.e.\ assuming either a flat overburden  or the Double Chooz hill profile.
The uncertainty due to rock composition is 3.5\% and comes from comparing our results for ``standard'' rock (density 2.65 g/cm$^3$) to those for Chooz rock (density 2.80 g/cm$^3$).
An overall systematic error of 6.1\% on mean muon energies takes into account in addition the numerical approximations introduced in the simulation and the uncertainty on primary muon flux.

The muon fluxes and mean energies at the various experimental sites are shown in Table \ref{tab:muenergy};
they are in good agreeement with the values quoted in \cite{DB}.

\begin{table}[tb]
\begin{center}
\caption{Muon flux and mean muon energy at the DC near,
Daya Bay (DB) and RENO experimental sites. 
\label{tab:muenergy}}
\begin{tabular}{lccccc}
\hline\hline
Detector     & depth & \multicolumn{2}{c}{$\Phi_\mu$ (m$^{-2}$s$^{-1}$)} & \multicolumn{2}{c}{$\langle E_\mu\rangle$ (GeV)} \\
             & (m.w.e.)  & quoted & calculated & quoted & calculated \\
\hline
RENO Near    & 120 & N/A   & $4.84 \pm 0.27$  & N/A    	& $33.3\pm2.0$ 	\\
DC Near      & 150 & N/A   & $3.12 \pm 0.17$ & N/A      & $39.7\pm2.4$  \\
DB EH1       & 250 & 1.27  & $1.08 \pm 0.06$ & 57  	& $58.5\pm3.6$ 	\\
DB EH2       & 265 & 0.95  & $0.95 \pm 0.05$ & 58  	& $61.0\pm3.7$ 	\\
RENO Far     & 450 & N/A   & $0.28 \pm 0.02$ & N/A    	& $89.3\pm5.4$ 	\\
DB EH3       & 860 & 0.056 & $0.05 \pm 0.01$ & 137 	& $139.8\pm8.5$ \\
\hline\hline 
\end{tabular} 
\end{center}
\end{table}

The rates of IBD candidates from fast neutrons and  $\beta$-$n$ isotopes were assumed to scale with depth ($h$) according to power laws \cite{Zatsepin, Wang}: $$ R_{n/\beta-n}(h) \propto  \Phi_\mu(h) \cdot \langle E_\mu(h)\rangle^\alpha\,.$$
Factors due to scintillator composition, summarized in Table~\ref{tab:ScintComp2}, were taken into account, and affect the results by no more than 3\%.
Background rates can depend on several other aspects of the experimental apparatus: acceptance, $\mu$ detection efficiency, neutron shielding type and thickness, selection cuts, etc.   
Thus, detailed use of these rates for other experiments requires corrections  to adapt from our detector to the detector of interest.

\begin{table}[t]
\begin{center}
\caption{Different liquid scintillator (LS) properties 
used for background rate scaling.
$M$ indicates the total mass and
$m_{LS}$ the molecular mass of the LS,
$N_{C/LS}$ and $N_{H/LS}$ are the number of carbon or hydrogen atoms 
per molecule of LS, $N_C$ ($N_H$) the total number of carbon (hydrogen) atoms 
in the detector target.
\label{tab:ScintComp2}}
\begin{tabular}{lccccccc}
\hline \hline
Experiment  & $M$      & $m_{LS}$ & $N_{C_{LS}}$ & $N_{H_{LS}}$  & $N_C$ & $N_H$ \\
            & (tons) & (g/mol)  &                   &  & $(10^{29})$ &$(10^{29})$  \\
\hline\hline
DC          & 8.24  & 178.33 & 12.67 & 24.65 & 3.53 &   6.75 \\
RENO         & 16.0  & 246.43 & 18    & 30 & 7.04 &  11.7 \\
Daya Bay     & 20.0  & 246.43 & 18    & 30 & 8.80 &  14.7 \\
KamLAND      & 913.4 & 160.31 & 11 & 22  & 385 & 767 \\
\hline \hline
\end{tabular} 
\end{center}
\end{table}

\begin{figure}[htb]
\includegraphics[width=\linewidth]{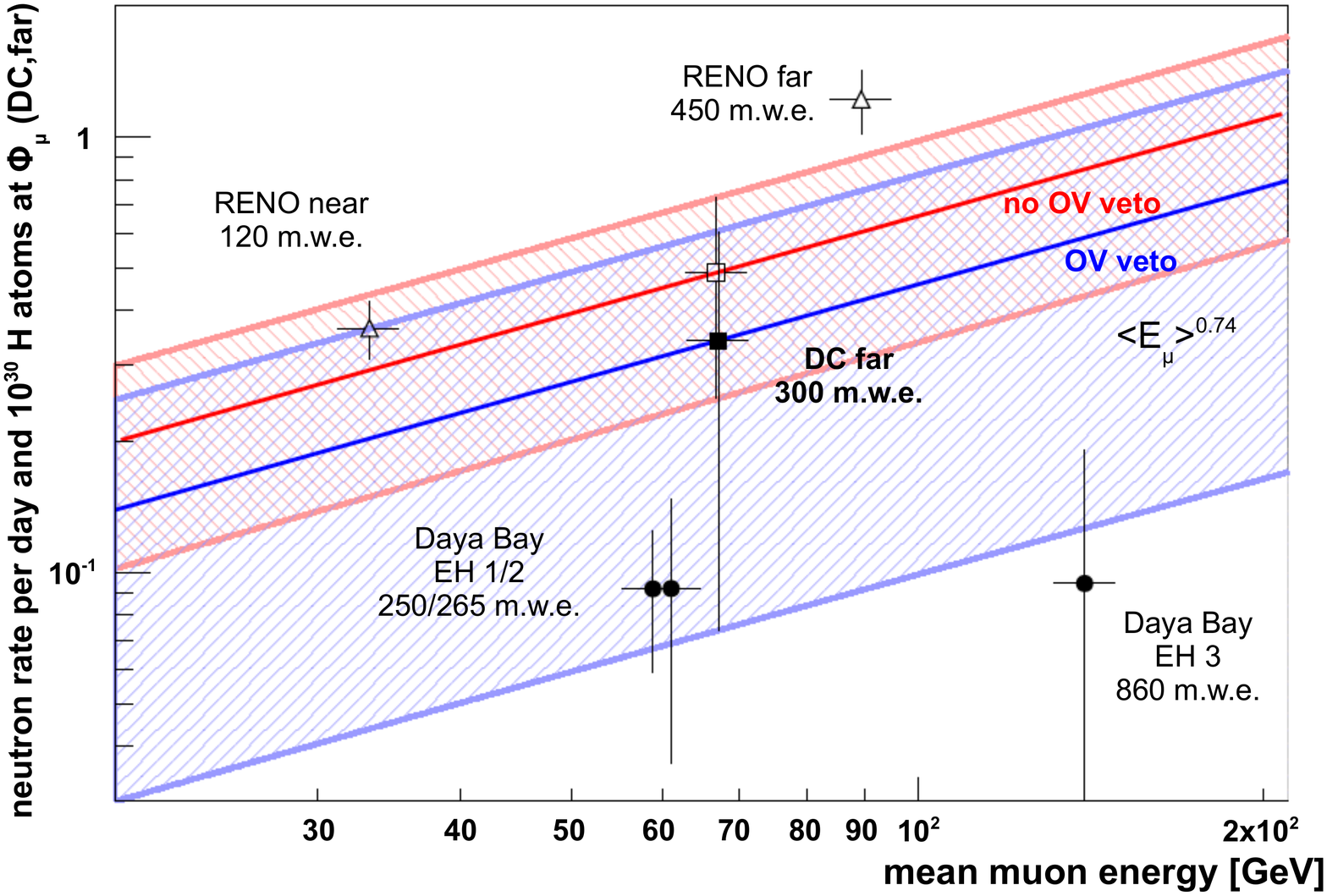} 
\caption{Scaling of DC fast-$n$ background rates and 
comparison with quoted values.
Empty (full) markers indicate quoted results
using a selection without (with) an external muon
veto; lines and shaded bands represent our scaling
of the DC measurements with their uncertainty.
Values were scaled by number of H atoms
and normalized to muon flux at DC far site.
\label{fig:rescn}}
\end{figure}

For fast-$n$, $\alpha=0.74$ is used, as estimated in~\cite{Zatsepin,Wang} from rates measured by several experiments at different depths.
The prompt signal in fast-$n$ background events arises from the recoil of a free proton in the target; 
for simplicity, we scale the rate to the number of hydrogen 
atoms in the target scintillators, assuming that interactions scale with detector volume, as is frequently done in the literature. 
The results are summarized in Table~\ref{tab:rescn} and compared to measured values~\cite{DB,RENO}, normalized to the muon flux at the DC far site, in Fig.~\ref{fig:rescn}.
The value quoted by RENO is obtained without 
a dedicated muon veto, 
and is thus comparable to our DCI result, while Daya Bay applies 
a water muon veto and is thus more similar to our DCII results.
The Daya Bay measurements are lower than our extrapolation, which could be due to the water surrounding their detectors.
For RENO, our extrapolation yields lower values than the measured ones.

\begin{table}[tb]
\caption{Fast-$n$ background rates measured at DC far and scaled to other depths.
\label{tab:rescn}}
\begin{tabular}{lccc}
\hline\hline
& & \multicolumn{2}{c}{\bf Fast-$n$ background rate}\\
Detector & depth & \multicolumn{2}{c}{\bf (day $\cdot 10^{30}$H)$^{-1}$} \\
 & (m.w.e.) & no OV veto & OV veto\\
\hline
RENO near    & 120 &  2.0 $\pm$ 1.0  &  1.4 $\pm$ 1.1  \\
DC near      & 150 & 1.44 $\pm$ 0.76 & 1.01 $\pm$ 0.82  \\
Daya Bay EH1 & 250 & 0.67 $\pm$ 0.33 & 0.46 $\pm$ 0.37  \\
Daya Bay EH2 & 265 & 0.60 $\pm$ 0.30 & 0.42 $\pm$ 0.33  \\
DC far       & 300 & 0.49 $\pm$ 0.24 & 0.34 $\pm$ 0.27  \\
RENO far     & 450 & 0.24 $\pm$ 0.12 & 0.16 $\pm$ 0.13  \\ 
Daya Bay EH3 & 860 & 0.06 $\pm$ 0.03 & 0.04 $\pm$ 0.03  \\ 
\hline\hline
\end{tabular}
\end{table}

For the scaling of {$\beta$-$n$} rates, the exponent $\alpha$ has never been measured experimentally. 
In~\cite{Hagner}, the combined rate of {$^9$Li} and {$^8$He} was measured at a single energy,
and the value $\alpha =$ 0.73$\pm$0.10 was used to extrapolate this rate to KamLAND and Borexino energies. 
In~\cite{KamLAND}, the value $\alpha =$ 0.801$\pm$0.026 is given for $\beta$-$n$ based on FLUKA simulations for various muon energies. A similar simulation, based on GEANT4, is described in~\cite{Zbiri}, where the resulting value for $\alpha$ is 1.06. 
To be conservative, we choose $\alpha =$ 0.84$\pm$0.22, ranging from the lower bound of~\cite{Hagner} to the result of~\cite{Zbiri}.

As cosmogenic isotope production scales with the number of target carbon atoms, rates are normalized to the total number of carbon 
atoms in the target scintillator.

\begin{figure}[htb]
\includegraphics[width=\linewidth]{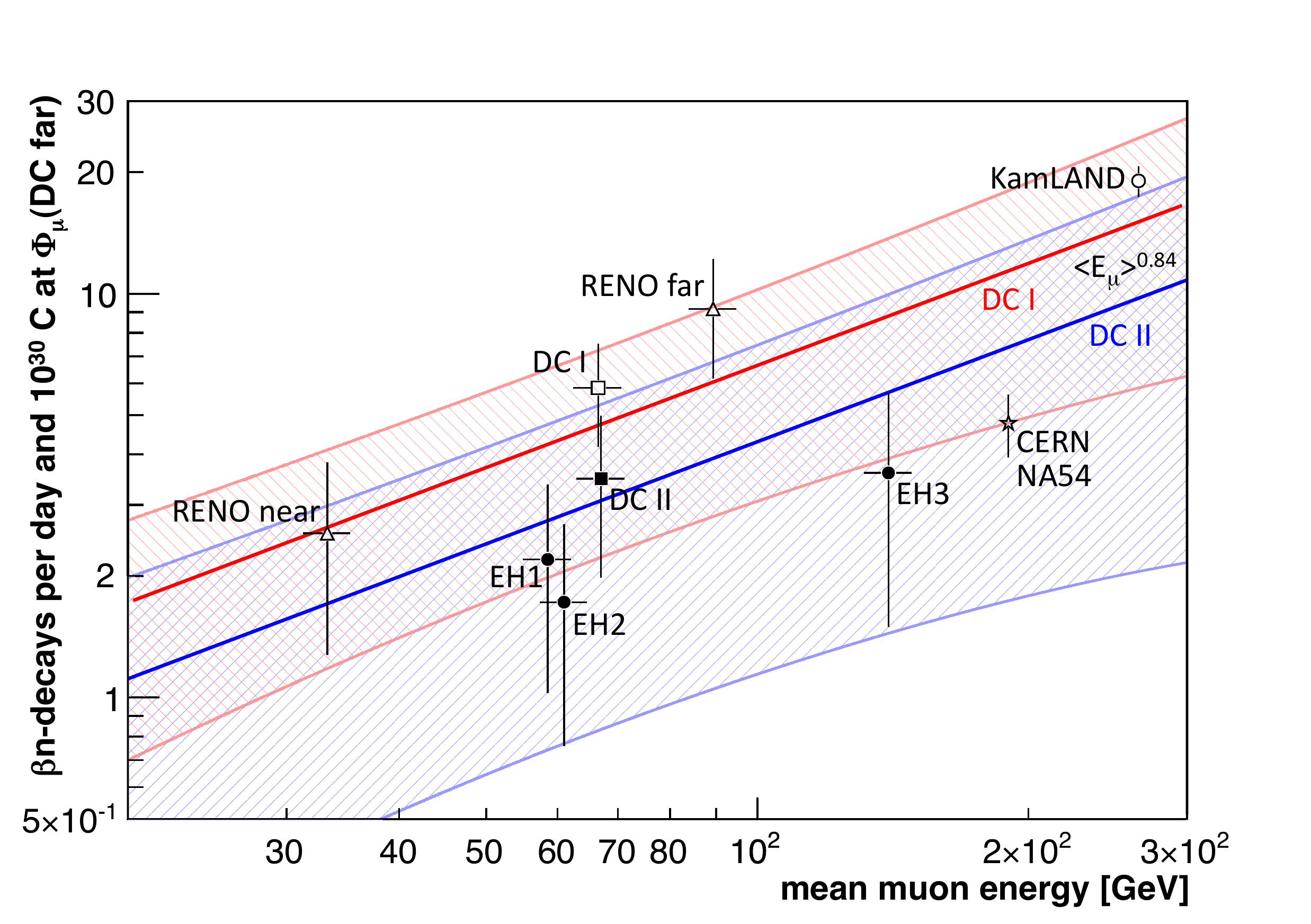} 
\caption{
Scaling of DC $\beta$-$n$ decay  rates and
comparison with quoted values.
Results were scaled by number of carbon atoms 
and normalized to muon 
flux at DC far site. 
Solid lines and shaded regions correspond to
 rate and scaling uncertainties in reactor-off analysis: 
DCI (red solid line) and open data
points compare the total $\beta$-$n$ rate, while DCII (blue solid line) 
and filled data points correspond to analyses with an extended veto 
following showering muons.
\label{fig:scaleLi}}
\end{figure}

Results for scaled  $\beta$-$n$ rates are shown in Table~\ref{tab:scaleLi} and compared to the measured values \cite{dc2012,DB,RENO}, normalized to the muon flux at the DC far site, in Fig.~\ref{fig:scaleLi}.
The DCII result is comparable to the Daya Bay value, where a veto of 1 s following showering muons has been applied, while the DCI result is comparable to the RENO one, with no specific $\beta$-$n$ background reduction.
No correction has been applied for the efficiency of the showering-$\mu$ veto.
Within the uncertainty of the measured $\beta$-$n$ rate, the scaled results agree.

\begin{table}[htb]
\caption{$\beta$-$n$ decay rates measured at DC far and scaled to other depths.\label{tab:scaleLi}}
\begin{tabular}{lccc}
\hline\hline
& & \multicolumn{2}{c}{\bf $\beta$-$n$-decay rate}\\
Detector & depth & \multicolumn{2}{c}{\bf (day $\cdot 10^{30}$C)$^{-1}$} \\
        & (m.w.e.) & DCI & DCII \\
\hline
RENO near    & 120 &   18 $\pm$ 10   & 11.7 $\pm$ 8.9 \\
DC near      & 150 & 13.5 $\pm$ 7.9  &  8.7 $\pm$ 6.7 \\
Daya Bay EH1 & 250 &  6.5 $\pm$ 3.5  &  4.2 $\pm$ 3.1 \\
Daya Bay EH2 & 265 &  5.9 $\pm$ 3.2  &  3.8 $\pm$ 2.8 \\
DC far       & 300 &  4.8 $\pm$ 2.6  &  3.1 $\pm$ 2.3 \\
RENO far     & 450 &  2.4 $\pm$ 1.3  &  1.5 $\pm$ 1.2 \\
Daya Bay EH3 & 860 & 0.63 $\pm$ 0.36 & 0.41 $\pm$ 0.31 \\
\hline\hline
\end{tabular}
\end{table}

In conclusion, we have reported a direct measurement of the cosmic-ray-induced background in the DC oscillation analysis using 7.53 days of data  with both reactors off. 
The identified candidates are well understood as due to accidentals,
$\beta$-$n$-emitting isotopes,  cosmic muons producing fast neutrons,
and  stopped muons that decay.   With the same cuts applied as
in the Double Chooz reactor-on oscillation analysis \cite{dc2012}, the total background including accidentals, cosmogenic $\beta$-$n$-emitting isotopes, fast neutrons from cosmic muons and stopped-$\mu$ decays  is 1.0$\pm$0.4 events/day.
The result is consistent with estimations in the DC oscillation analysis.
The results have been scaled to depths of interest to the Daya Bay and RENO reactor-based neutrino oscillation experiments.   

\section*{Acknowledgments}

We are grateful to Vitaly Kudryavtsev for providing and supporting the MUSIC and
MUSUN muon transport codes.
We thank the French electricity company EDF; the
European fund FEDER; the R\'egion de Champagne Ardenne; 
the D\'epartement des Ardennes; and the Communaut\'e des Communes
Ardennes Rives de Meuse. We acknowledge
the support of the CEA,  CNRS/IN2P3, CCIN2P3 and 
LabEx UnivEarthS in France; the Ministry of Education, Culture,
Sports, Science and Technology of Japan (MEXT) and
the Japan Society for the Promotion of Science (JSPS);
the Department of Energy and the National Science
Foundation of the United States; the Ministerio de Ciencia
e Innovaci´on (MICINN) of Spain; the Max Planck
Gesellschaft, the Deutsche Forschungsgemeinschaft
DFG (SBH WI 2152), the Transregional Collaborative
Research Center TR27, the Excellence Cluster ``Origin 
and Structure of the Universe,'' the Maier-Leibnitz-Laboratorium 
Garching and the SFB676 in Germany; the Russian Academy of Science,
the Kurchatov Institute and RFBR (the Russian
Foundation for Basic Research); and the Brazilian Ministry
of Science, Technology and Innovation (MCTI), the Financiadora
de Estudos e Projetos (FINEP), the Conselho Nacional de 
Desenvolvimento Cient\'{i}fico e Tecnol\'{o}gico 
(CNPq), the S˜ao Paulo Research Foundation (FAPESP),
the Brazilian Network for High Energy Physics (RENAFAE)
in Brazil.

\end{document}